\documentclass[prb,twocolumn,showpacs,preprintnumbers,amsmath,amssymb]{revtex4}

\usepackage{graphicx}
\usepackage{dcolumn}
\usepackage{bm}
\usepackage{amsmath}

\usepackage[dvips]{color}
\usepackage{ulem}

\begin{document}

\title
{
Correlated electron transport through double quantum dots coupled to
 normal and superconducting leads
}

\author{
Yoichi Tanaka$^1$, Norio Kawakami$^2$, and Akira Oguri$^{3}$
}

\affiliation{
$^{1}$Condensed Matter Theory Laboratory, RIKEN, 
Saitama 351-0198, Japan\\
$^{2}$Department of Physics, Kyoto University, Kyoto 606-8502, Japan
\\
$^{3}$Department of Physics, Osaka City University, Osaka 558-8585, 
Japan
}%

\date{\today}

\begin{abstract}
We study Andreev transport through double quantum dots connected in series normal and superconducting (SC) leads, using the numerical renormalization group. 
The ground state of this system 
shows a crossover between a local Cooper-pairing singlet state and 
a Kondo singlet state, which is caused by 
the competition between the Coulomb interaction and the SC proximity.
We show that the ground-state properties reflect this 
crossover especially for small values of the inter-dot coupling $t$, 
 while in the opposite case, for large $t$, 
 another singlet with an inter-dot character becomes dominant.
We find that the conductance for the local SC singlet state has 
a peak with the unitary-limit value $4e^2/h$. 
In contrast, the Andreev reflection is suppressed 
in the Kondo regime by the Coulomb interaction.
Furthermore, the conductance has two successive peaks   
in the transient region of the crossover.  
It is further elucidated that 
the gate voltage gives a different variation into the crossover. 
Specifically, as the energy level of the dot 
that is coupled to the normal lead varies, 
the Kondo screening cloud is deformed to a long-range singlet bond.

\end{abstract}


\maketitle

\begin{figure}[b]
\includegraphics[scale=0.7]{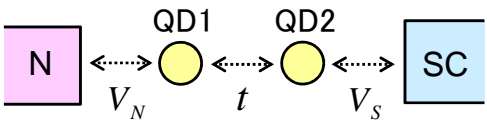}
\caption{
(Color online) 
Quantum dots (QD's) coupled to normal (N) and superconducting (SC)
leads in series. 
}
\label{modelNsDQDS}
\end{figure}

\section{Introduction}
A quantum dot (QD) \cite{Kouwen} is a prototype of nanoscale systems,
and has offered unprecedented opportunities
to uncover correlation effects on quantum transport.
In particular, the Kondo effect in a QD system gives rise to 
remarkable properties in electron transport, 
\cite{Gold,Cronen,Glaz,Ng,MW}
which has encouraged further studies in this field.
With recent experimental advancements, it becomes possible to examine 
the Kondo physics in a variety of systems, 
such as an Aharonov-Bohm (AB) ring with 
a single QD and double quantum dots (DQD).

A QD coupled to superconducting leads 
is also an intriguing system, which can bring about the competition between 
superconductivity and the Kondo effect.
In fact, such a competition has recently been observed  
in carbon nanotube and semiconductor QD systems.
\cite{Buit:2002,Eich:2007,Sand:2007,Buiz:2007,Grov:2007,SDQDS_exp}
 Furthermore the Andreev scattering at a junction of 
a normal metal (N)  and superconducting (SC) leads,
is also a fascinating phenomenon.
Particularly, in the QD's coupled to the normal and SC leads  
a short-range Cooper pair penetrates 
into the QD as a linear combination of 
the empty and doubly occupied states, so that the SC proximity 
becomes sensitive to the Coulomb repulsion.
The interplay between the Andreev scattering 
and the Kondo effect takes place in such a situation, 
and has been studied intensively for a single QD, 
theoretically\cite{Fazio,Schwab,Cho,Clerk,Sun,Cuevas,Avishai1,Aono,Krawiec,Splett,Doma,TanakaNQDS} and experimentally.\cite{Graber,Deacon}
We have studied the ground-state properties of 
a single dot Andreev-Kondo system, 
and have confirmed in a previous work\cite{TanakaNQDS} that 
the zero-temperature conductance shows a maximum at 
a crossover region between the Kondo singlet state 
and the local Cooper-pairing singlet state.
The local Cooper pairing is caused by the SC proximity effect, and 
consists of a linear combination of the empty and doubly occupied states.
We shall refer to this singlet state as the local SC singlet state for short.

The Andreev transport has also been explored 
further in other nanosystems,  e.g., 
an AB ring with a single QD,\cite{Avishai2}
a DQD system \cite{Tanaka:NTDQDS} and molecular wires. \cite{Korma}
Furthermore, the Josephson current in a serial DQD system,
where both leads connected to dots are in SC states,
 has also been investigated in recent years.
\cite{Berg:2006,Lopez:2007}
It is clarified that the inter-dot coupling plays a crucial role on 
the Josephson effect. 
In fact, one of the typical features of the DQD systems 
is that another type of a singlet ground state, 
which is referred to as an inter-dot singlet hereafter,
appears and plays an important role on the low-energy properties.
To our knowledge, however, there have been few studies of the Andreev transport through the DQD system so far,
while a lot of studies have been done for the normal transport 
in the DQD system.
 The underlying Andreev-Kondo physics in the DQD system 
is still less understood, 
and is needed to be clarified precisely.
Moreover, the DQD system coupled to normal and SC leads can be fabricated, 
\cite{Hofstetter,Herrmann}
and is studied from the viewpoint of spin entanglement.
\cite{Recher}
These developments are also spurring further research 
of the DQD system coupled to normal and SC leads.


In the present paper, 
we theoretically study the transport through the DQD
coupled in series to normal and SC leads,
as shown in Fig.\ \ref{modelNsDQDS}.
The SC proximity into the DQD 
is affected more by the Coulomb interaction $U_2$
at the QD2, which is located adjacent to the SC lead, 
than the interaction $U_1$ at the QD1, 
sitting away from the SC lead (see Fig.\ \ref{modelNsDQDS}).
Such a spatial variation gives a variety, 
which cannot be realized in the single dot,
into the Andreev-Kondo physics in the DQD system.
Particularly, we focus on the crossover between the ground states
which can be classified into the local SC singlet,
 the Kondo singlet and the inter-dot singlet states,
 and clarify how the crossover affects the transport properties.
 We first of all demonstrate that in the electron-hole symmetric case 
this system can be mapped onto the two impurity 
Anderson model coupled to normal leads. 
Then, we calculate the conductance due to the Andreev reflection
in a wide parameter range of the inter-dot coupling, the Coulomb interaction, 
 and the energy level of the quantized states in the dots, 
using the numerical renormalization group (NRG) method\cite{Wilkins,Hewson1}
in the limit of a large SC gap $\Delta \to \infty$. 
In the case that the Coulomb interaction $U_2$ at the QD2 is small,
the short-range Cooper pair can penetrate into the QD2,
 and the conductance takes the unitary-limit value $4e^2/h$. 
In contrast, for large $U_2$ the SC proximity to the QD2 is 
suppressed, and the conductance does not reach the unitary-limit value.
It reflects a significant change of
 the SC correlation penetrating into the DQD,
which we have deduced from the behavior of 
the renormalized parameters for the Bogoliubov particles.
Furthermore, we find that the conductance has
two peaks with the unitary-limit value
near the crossover between the Kondo singlet state and 
the local SC singlet state.
 The contrast between the two singlet states is pronounced 
 for small inter-dot coupling $t$. In the opposite limit, for large $t$, 
the ground state becomes another singlet with an inter-dot character.
We examine also the gate voltage dependence of the conductance, 
varying the energy level of each dot separately. 
The crossover between these singlet states 
takes place as the energy levels are varied, 
and near the crossover point the conductance shows a peak.
It is further found that a different type of Kondo 
singlet with a long-range singlet bond emerges, 
as the energy level of the QD1 which is located at the normal-lead 
side varies.


This paper is organized as follows. 
In the next section, we introduce the model and 
give a brief outline of the Bogoliubov transformation.
Then in Sec. \ref{sec:result}, we show the numerical results and discuss 
the Andreev transport, focusing on the interplay among 
the Kondo effect, the SC correlation and the inter-dot coupling.
Moreover, the results for the energy-level dependence of the conductance
are presented.
A brief summary is given in the last section.

\section{Model and Formulation} \label{sec:model}

\subsection{Model}
The Hamiltonian of a DQD coupled in series to normal (N) and
superconducting (SC) leads is given by
\begin{eqnarray}
H &=& H_{DQD} + H_S + H_N + H_{T,S} + H_{T,N},
\label{Hami_seri}
\end{eqnarray}
where $H_{DQD}$ and $H_{S(N)}$ represent the DQD part and
the SC (normal) lead part, respectively. 
$H_{T,S(N)}$ is the mixing term between the QD and the SC (normal) lead.
The explicit form of each part reads 
\begin{align}
&H_{DQD}
=
\sum_{i=1,2}
\left\{
\xi_{i} \left(n_{d,i}-1\right)
+\frac{U_i}{2}\left(n_{d,i}-1\right)^2
\right\}
\nonumber\\
& \qquad \qquad  
+t\,\sum_{\sigma}\left(d_{1\sigma}^{\dag}d_{2\sigma}^{}+
d_{2\sigma}^{\dag}d_{1\sigma}^{}
\right),
\nonumber\\
&H_S \, 
=
\sum_{k,\sigma}\varepsilon _{S,k}
c_{S,k\sigma}^\dag c_{S,k\sigma}^{}
-\sum_{k}\left(\Delta\, 
c_{S,k\uparrow}^\dag \,c_{S,-k\downarrow}^\dag 
+ \textrm{H.c.}\right),
\nonumber\\
& 
H_N 
= \sum_{k,\sigma}\varepsilon _{N,k}
c_{N,k\sigma}^\dag c_{N,k\sigma}^{},
\nonumber \\
&H_{T,N} = \sum_{k,\sigma} \frac{V_{N}}{\sqrt{\mathcal{N}}}
\left(c_{N,k\sigma}^\dag d_{1\sigma}^{} + 
 d_{1\sigma}^{\dag}c_{N,k\sigma}^{} \right),
\nonumber \\
&H_{T,S} = \sum_{k,\sigma} \frac{V_{S}}{\sqrt{\mathcal{N}}}
\left(c_{S,k\sigma}^\dag d_{2\sigma}^{} + 
 d_{2\sigma}^{\dag}c_{S,k\sigma}^{} \right).
\label{Hami_seri_part}
\end{align}
Here, $\xi_{i}\equiv\varepsilon_{i}+U_i/2$ for $i=1,2$. 
The operator $d^{\dag}_{1(2)\sigma}$ creates an electron with 
energy $\varepsilon _{1(2)}$ and spin $\sigma$ at the QD1(QD2). 
$U_{1(2)}$ is the Coulomb interaction, and 
$t$ is
the inter-dot coupling, 
and 
$n_{d,i}=\sum_{\sigma}d^{\dag}_{i\sigma}d^{}_{i\sigma}$.
$c_{S(N),k\sigma}^\dag$ denotes the creation operator
of an electron with the energy $\varepsilon _{S(N),k}$
in the SC (normal) lead. 
$V_{S/N}$ is the tunneling matrix element between the QD2/QD1 and 
the SC/normal lead, and $\Delta$ is an $s$-wave BCS gap. 
We assume that $\Gamma_{S/N}(\varepsilon) \equiv 
\pi V_{S/N}^{2} \sum_k \delta(\varepsilon-\varepsilon_{k})/\mathcal{N}$ 
is a constant independent of the energy $\varepsilon$, 
where $\mathcal{N}$ is the number of the states in each lead.

To be specific,
we concentrate on a large SC gap limit $\Delta \to \infty$ 
in the  present paper. 
In this limit the quasi-particle excitations 
in the continuum energy region above the SC gap are projected out. 
Nevertheless, 
the Andreev reflection takes place inside the SC gap,
and the essential physics of the low-energy transport 
is preserved still in the large 
gap limit.\cite{condition}
In this case, the starting Hamiltonian $H$ can be mapped exactly 
onto a single-channel model 
for which the NRG approach works more 
efficiently,\cite{TanakaNQDS,Affleck,Oguri}
\begin{align}
H^\mathrm{eff} =& H^\mathrm{eff}_S + H_{DQD} + H_N + H_{T,N}\;, 
\label{Hami_seri_eff}\\
H^\mathrm{eff}_S = & \,  
-\Delta_{d2} 
\left(d^{\dag}_{2\uparrow }d^{\dag}_{2\downarrow }+\textrm{H.c.}
\right) \;, 
\label{Hamieff_S} \\
\Delta_{d2} \equiv & \  \Gamma_S \;.
\label{Delta_d2}
\end{align}
Here, an additional term $H^\mathrm{eff}_S$ appears 
instead of the SC lead $H_S + H_{T,S}$. 
The SC proximity effect becomes static in the large gap limit, 
and is described by the pair potential 
in the QD2,  $\Delta_{d2} \equiv \Gamma_S$. 
This term breaks the charge conservation of the electrons, 
and causes the Andreev reflection.

\subsection{Bogoliubov transformation in the $\xi_1=0$ case}
\label{subsec:bogo}

The system described by $H^\mathrm{eff}$ has 
a conserved charge in the case 
that the value of the energy level of the QD1 
satisfies the condition $\xi_{1}=0$. 
 Then the Hamiltonian $H^\mathrm{eff}$ 
can be transformed into an asymmetric two-impurity 
Anderson model for the Bogoliubov particles, 
the total number of which 
is conserved.\cite{TanakaNQDS,Tanaka:NTDQDS,Yoshihide}
This is due to the rotational symmetry in the Nambu pseudo-spin space, 
in which $\xi_i$ and $\Delta_{d2}$ can be regarded, 
respectively, as the $z$ and $x$ components 
of an external field $\vec{\bm{\eta}}$ which 
couples to the pseudo spin.\cite{Yoshihide}
Specifically for $\xi_{1}=0$, 
the external field becomes finite only in the QD2,
and its contribution to the energy is given by 
$\vec{\bm{\eta}}\cdot\vec{\bm{\tau}} 
= -\Delta_{d2} \bm{\tau}_x + \xi_2 \bm{\tau}_z$,
where $\bm{\tau}_j$ for $j=x,y,z$ is the Pauli matrix 
in the Nambu representation.
Thus the system has a uniaxial symmetry along the direction 
of the local external field  $\vec{\bm{\eta}}$ in the pseudo-spin space.

\begin{figure}[t]
\includegraphics[scale=0.53]{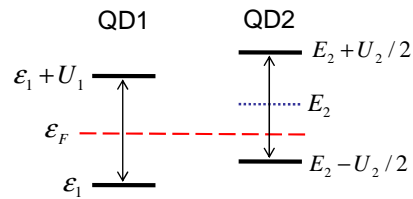}
\caption{
(Color online) 
Schematic of the energy level profile at the DQD part for
 the system after the Bogoliubov transformation.
The (red) dashed line denotes the Fermi energy of the leads 
($\varepsilon_F \equiv 0$),
and the (blue) dotted line indicates $E_2$, which
is located at the center between two energy levels, $E_2-U_2/2$ 
and $E_2+U_2/2$.
}
\label{sketchDQD}
\end{figure}

In order to use these symmetry properties, 
we rewrite the normal lead part of the Hamiltonian, 
$H_N$ and $H_{T,N}$, in a tight-binding form
\begin{align}
&H_{N} =
\sum_{n=0}^{\infty} \sum_{\sigma} 
t_{N,n}^{\phantom{0}} 
\left(
f^{\dagger}_{n+1\sigma}
f^{\phantom{\dagger}}_{n\sigma}
+
f^{\dagger}_{n\sigma} 
f^{\phantom{\dagger}}_{n+1\sigma}
\right) , 
\label{eq:HN-ti}
\\
&H_{T,N} =
\sum_{\sigma} 
V_{N}^{\phantom 0}  
\left(\,
 f^{\dagger}_{0 \sigma} \, d^{\phantom{\dagger}}_{1\sigma} 
\, + \,  
 d^{\dagger}_{1\sigma} \,
 f^{\phantom{\dagger}}_{0 \sigma}  
\, \right) .
\label{eq:Hmix-ti}
\end{align}
Here, $f^{}_{0 \sigma}
= \sum_{k} c_{N,k\sigma}^{}/\sqrt{\mathcal{N}}$.
Note that no approximation has been made to obtain Eq.\ \eqref{eq:HN-ti},  
and the hopping matrix element $t_{N,n}$ can be  
generated from $\varepsilon _{N,k}$ 
via the Householder transformation.\cite{Hewson1}
The tight-binding form reveals  
the electron-hole symmetry of $H_{N}$ explicitly.
In the case $\xi_{1} =0$, the Hamiltonian can be simplified by
the Bogoliubov transform, 
taking the direction of the local external field at the QD2 as 
a new quantization axis in the pseudo-spin space, 
\begin{align}
&\left[
 \begin{array}{c}
  \gamma_{n\uparrow}^{\phantom{\dagger}} \\
  (-1)^{n} \gamma_{n\downarrow}^{\dagger}
 \end{array}
\right]
=
\left[ 
        \begin{array}{cc}
          u  &  -v \\
          v  &  \ u \rule{0cm}{0.5cm}
        \end{array} 
\right]
\left[
 \begin{array}{c}
  f_{n\uparrow}^{\phantom{\dagger}} \\
  (-1)^{n} f_{n\downarrow}^{\dagger}
 \end{array}
\right]\;, 
\label{eq:Bogo_B}
\\
\nonumber \\
&
   u=\sqrt{\frac{1}{2}\left(1+\frac{\xi_{2}}{E_{2}}\right)}, \quad
   v=\sqrt{\frac{1}{2}\left(1-\frac{\xi_{2}}{E_{2}}\right)} 
\label{eq:Bogo_factor_B} 
\;,\\
\nonumber \\ 
&E_{2} \equiv \sqrt{\xi_{2}^2+\Delta_{d2}^{2}}
\label{eq:E_2}
\;.
\end{align}
The transformation is carried out for the whole cites, 
$n \geq -2$, including the DQD part 
for which we use a notation $f_{-i\sigma} = d_{i\sigma}$ for $i=1,2$.
Similarly, the hopping matrix elements $\,t_{N,n}$ for $n<0$ 
are defined to be  $t_{N,-1} \equiv V_N$ and $t_{N,-2}\equiv t$.
Then 
the effective Hamiltonian $H^\mathrm{eff}$ can be expressed in the form  
\begin{align}
&\!\!\!
H^\mathrm{eff}=
E_{2} 
\left(\widehat{n}_{\gamma, -2} -1
\right)+\sum_{i=1,2} \frac{U_{i}}{2}\left(\widehat{n}_{\gamma,-i}-1\right)^2
\nonumber\\
&
\qquad 
+\sum_{n=-2}^{\infty} \sum_{\sigma}
t_{N,n}^{\phantom{0}} 
\left(
\gamma_{n+1\sigma}^{\dagger}\gamma_{n\sigma}^{\phantom{\dagger}}
+\textrm{H.c.}
\right) .
\label{eq:after-Bogo}
\end{align}
Here,  $\widehat{n}_{\gamma,-i} = \sum_{\sigma} 
\gamma_{-i\sigma}^{\dagger}\gamma_{-i\sigma}^{\phantom{\dagger}}$
is the number of Bogoliubov particles at the QD$i$ for $i=1,2$.
In this representation,  the value of $E_{2}-U_2/2$
corresponds to an energy level
for the Bogoliubov particles in the QD2.  
This correspondence is also illustrated schematically in Fig.\ \ref{sketchDQD}.
 For instance in the atomic limit $t_{N,n}^{\phantom{0}} \to 0 $,
the QD2 tends to be occupied by 
 a single Bogoliubov particle for $E_{2}-U_2/2<0$, 
while the QD2 tends to be empty for $E_{2}-U_2/2>0$.
The equation \eqref{eq:after-Bogo} 
clearly shows that the total number of Bogoliubov particles 
$\widehat{{\mathcal Q}}_\mathrm{tot}$ is conserved,
\begin{align}
\widehat{{\mathcal Q}}_\mathrm{tot} \equiv \sum_{j=-2}^{\infty} 
\widehat{n}_{\gamma,j}
\;.
\end{align}
It should also be noted that the Friedel sum rule holds 
for the number of the Bogoliubov particles
in the DQD, \cite{Langer,Sum:Tanaka}
\begin{align}
{\mathcal Q} \equiv 
\sum_{i=1,2} 
\langle \widehat{n}_{\gamma, -i} \rangle 
\  = \  \frac{2\,\varphi}{\pi} \;.
\label{MBogo}
\end{align}
Here, $\varphi$ is the phase shift of the Bogoliubov particles,
defined in the appendix \ref{sec:Deri-con}.
The phase shift $\varphi$ can be deduced from 
the low-lying eigenvalues of discretized Hamiltonian 
 of the NRG at the fixed point.\cite{Hewson2}

Furthermore, one can deduce the following quantities
from the value of ${\mathcal Q}$,
via the inverse transformation of Eq. \eqref{eq:Bogo_B}, 
\begin{align}
K 
&\equiv \,
\sum_{i=1,2} (-1)^{i}\, \langle \kappa_{d,i} \rangle
\ = -\frac{\Delta_{d2}}{E_{2}} ({\mathcal Q}-2),
\label{eq:K_sDQD}
\\
M &\equiv \,
\sum_{i=1,2} \bigl(\langle n_{d,i} \rangle -1\bigr)
\ =
\frac{\xi_{2}}{E_{2}} ({\mathcal Q}-2)        \;.
\label{eq:M_sDQD}
\end{align}
Here, 
$\kappa_{d,i} \equiv \, d_{i\uparrow }^\dag d_{i\downarrow}^\dag
 + d_{i\downarrow}^{} d_{i\uparrow}^{}$.
The SC correlation $K$ and 
the number of the original electrons $M$ in the DQD 
correspond, respectively, 
to the average value of the $x$ and $z$ components 
of the induced pseudo-spin moment.


In the case of $\xi_{1}=0$, the linear conductance $G$ at zero temperature
can also be expressed in terms of the local charge ${\mathcal Q}$, or
the phase shift $\varphi$, of the Bogoliubov particles 
(see also the appendix \ref{sec:Deri-con}),
\begin{align}
G
\, = \, 
\frac{4e^2}{h}
\left(
\frac
{\Delta_{d2}}{E_{2}}
\right)^2
\sin^2 \left(\pi {\mathcal Q} \right) .
\label{eq:Condu}
\end{align}

\subsection{Conductance for $\xi_1 \neq 0$}

For $\xi_{1} \neq 0$, however, 
the uniaxial symmetry in the pseudo-spin space is broken, 
and the system no longer has a conserved charge such as 
the total number of the Bogoliubov particles 
$\widehat{{\mathcal Q}}_\mathrm{tot}$.
 We calculate the conductance 
for these cases, using the Kubo formula,\cite{Izumida}
\begin{align}
&G \,=\, 
\lim_{\omega\to 0}\sum_{n} \frac{\pi\hbar}{\varepsilon_{n}}
\left| \langle {\rm GS}|J_N|n \rangle \right|^2
\delta (\hbar\omega-\varepsilon_{n}) \;,
\label{conKubo} \\
&J_N \,=\, \frac{ie}{\hbar}
\sum_{k,\sigma} \frac{V_{N}}{\sqrt{\mathcal{N}}} \, 
\left( d_{1\sigma}^{\dag} c_{N,k\sigma}^{} -
 c_{N,k\sigma}^{\dag} d_{1\sigma}^{}  \right).
\label{current}
\end{align}
Here, $\varepsilon_{n}$ and $|n \rangle$ denote 
an eigenvalue and the corresponding eigenstate of $H^\mathrm{eff}$,
 $|{\rm GS} \rangle$ is the ground state, 
and $J_N$ is the current which flows from the normal lead to the QD1.

%
\section{Numerical Results} \label{sec:result}
%

In this section, we provide the NRG results for 
the ground-state properties of 
the serial DQD coupled to normal and SC leads.
The calculations have been carried out, 
taking the tunneling matrix 
elements $t_{N,n}$ for $n \ge 0$ and $V_{N}$ in the forms
\begin{align}
&\!\!\!\!
t_{N,n} = D\, \frac{1+1/\Lambda}{2}
{ 1-1/\Lambda^{n+1}  
\over  \sqrt{1-1/\Lambda^{2n+1}}  \sqrt{1-1/\Lambda^{2n+3}} 
} \;,
\label{eq:tn}
\\
&\!\!
V_{N}=\sqrt{ \frac{2\,\Gamma_N D\,A_{\Lambda}}{\pi} },
\quad \ 
A_{\Lambda}=\frac{1}{2} 
\left(\, {1+1/\Lambda \over 1-1/\Lambda }\,\right)
\log \Lambda ,
\label{eq:t0}
\end{align}
where $D$ is the half-width of the conduction band.
We have kept lowest 1000 states in each NRG step, 
and have chosen the discretization parameter to be $\Lambda=3.0$.

\subsection{Dependence on the inter-dot coupling $\,t$}
\label{subsec:t-dep}
\begin{figure}[t]
\begin{center}
\includegraphics[trim=0mm 0mm 0mm 0mm, scale=0.75]{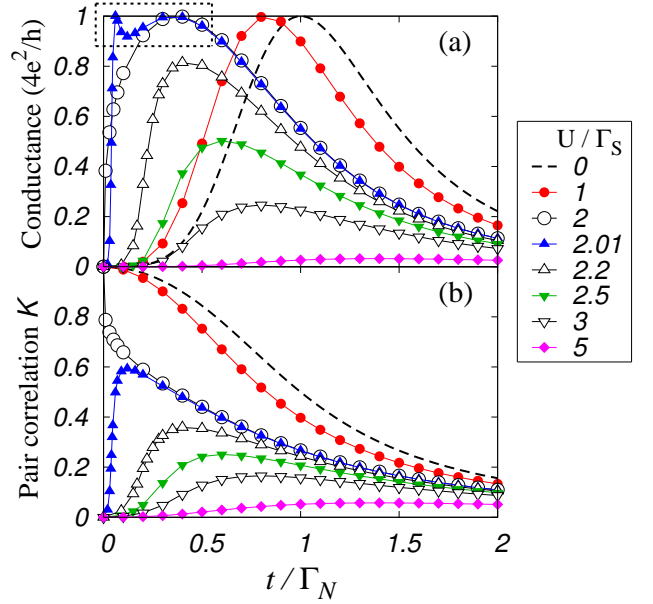}
\end{center}
\vspace{-5mm}
\caption{
(Color online) 
(a) Conductance $G$ and
(b) the pair correlations $K$
as a function of $t/\Gamma_N$ for  
$U_1=U_2=U$ and $\xi_1=\xi_2=0$. The hybridization energy scale is 
chosen to be $\Gamma_N=\Gamma_S$, and  
the SC proximity appears through $\Delta_{d2} \equiv \Gamma_S$.
The NRG calculations have been carried out for 
$\Lambda=3.0$ and $\Gamma_N/D=1.0\times 10^{-3}$.
}
\label{cont}
\end{figure}
\begin{figure}[t]
\begin{center}
\includegraphics[trim=0mm 0mm 0mm 0mm, scale=0.6]{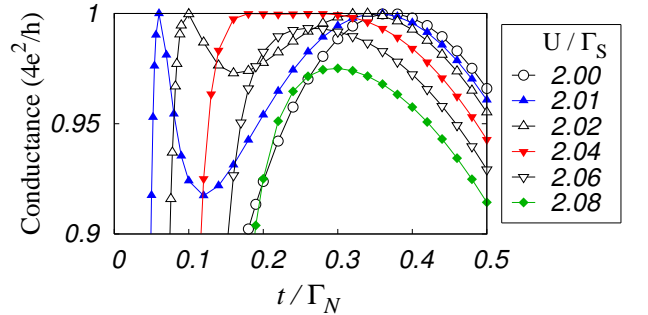}
\end{center}
\vspace{-5mm}
\caption{
(Color online) 
Conductance $G$ as a function of $t/\Gamma_N$ near $U/\Gamma_S=2.0$,
where the other parameters are the same as in Fig. \ref{cont}.
This figure corresponds to the enclosed area 
drawn by the dotted line in Fig. \ref{cont}.
}
\label{contU2}
\end{figure}

We first consider the case $U_{1}=U_{2}=U$, 
and choose the values of energy levels to be $\xi_1=\xi_2=0$.
In this case the energy level for the Bogoliubov particles in the QD2 
is given by $E_2 =\Delta_{d2}$ ($\equiv \Gamma_S$), and 
the conductance is determined by the local charge ${\mathcal Q}$
through Eq.\ \eqref{eq:Condu}.
The NRG results of the conductance are shown in Fig.\ \ref{cont} 
as a function of the inter-dot coupling $\,t$  
for several values of $U$,
for a fixed  hybridization strength $\Gamma_N=\Gamma_S$.
For weak interactions $U/\Gamma_S\le 2.0$, 
the conductance peak reaches the unitary-limit value $4e^2/h$,
and the peak position shifts towards the small $t$ side. 
In contrast, for $U/\Gamma_S \gtrsim 2.0$, 
the conductance peak does not reach the unitary-limit value. 
Note that the onsite level for the Bogoliubov particle 
in the present case is given by $E_2-U_2/2=\Gamma_S-U_2/2$.
Thus, for $\Gamma_S<U/2$, the Kondo effect takes place  
at the singly occupied QD2, 
as we will check out further in the next section.
We can see that there is a fine structure near $U/\Gamma_S=2.0$ 
for small values of $t$.
Figure \ref{contU2} is an enlarged picture 
of the conductance for the small $t$ region, which 
is marked with the dotted line in Fig.\ \ref{cont}(a). 
The conductance for $U/\Gamma_S=2.01$ 
has two unitary-limit peaks.
These two peaks get close to each other for $U/\Gamma_S=2.02$,  
and the conductance shows a plateau for $U/\Gamma_S=2.04$ 
at $0.16 \lesssim t/\Gamma_N \lesssim 0.3$.
Then, as $U$ increases further, the maximum value of the conductance 
does not reach the unitary-limit value.

In the present case, $\xi_{1}=\xi_{2}=0$, 
the SC pair correlation $K$ defined in Eq.~\eqref{eq:K_sDQD} 
and  the conductance $G$ in Eq. \eqref{eq:Condu}
can be expressed in the simplified forms, 
\begin{align}
K &\, = \, \frac{2}{\pi} (\pi-\varphi) ,
\qquad \quad 
G
\,=\,
\frac{4e^2}{h} 
\sin^2 2\varphi  \;.
\label{Kandphase}
\end{align}
Furthermore, the number of the real electrons takes 
a value of $M=0$ at half-filling. 
Note that the conductance takes the unitary-limit value $4e^2/h$ 
for $K=1/2$, which corresponds to $\varphi=3\pi/4$ and $\mathcal{Q}=3/2$. 
In Fig.~\ref{cont}(b),
the NRG results of $K$ are plotted as a function of $t/\Gamma_N$ 
for several values of $U/\Gamma_S$.  
We can see that the behavior of the pair correlation $K$ also changes 
significantly 
at the Coulomb interaction of the value of $U=2\Gamma_S$.
In the large $t$ limit,
two Bogoliubov particles occupy a bonding 
orbital consisting of the QD1 and QD2, 
so that $\mathcal{Q} \to 2.0$ and $K \to 0.0$ at $t/\Gamma_N \gg 1.0$. 
For weak interaction $U \le 2 \Gamma_S$, 
$K$ increases monotonically as $t$ decreases, 
and approaches the value of $K=1.0$, 
which corresponds to $\mathcal{Q}=1.0$, in the limit of $t \to 0$.
In contrast, for large interactions $U > 2\Gamma_S$,
 $K$ shows a peak  at an intermediate value of $t$, 
and then tends to $K=0.0$ as $t$ decreases.
The conductance has two peaks 
in the case that the maximum value of $K$ becomes $K>1/2$.
The behavior at $U < 2\Gamma_S$ 
is determined by the Andreev scattering due to the 
finite SC pair potential $\Delta_{d2} =\Gamma_S$, 
while for large $U > 2\Gamma_S$ 
the behavior is determined by the Kondo effect.

It seems to be meaningful 
to compare these results with the normal transport 
through the serial DQD coupled to two normal leads (N-DQD-N),
\cite{Lang,George,Rosa,Eto,Eto2,Izumida:sDQD,Sakano,Mrav,Jeong}
particularly with the NRG results by Izumida {\it et al}.\cite{Izumida:sDQD}
In the normal case, the conductance peak shifts towards
the small $t$ side as the Coulomb interaction $U$ increases, 
and  this behavior is similar to the one we observed for the 
Andreev transport in Fig. \ref{cont}(a).
However,  in the N-DQD-N case, the peak height of the conductance reaches 
the unitary-limit value $2e^2/h$ even in the large $U$ region.
Therefore, the suppression of the conductance for large $U$
and the two-peak structure observed for small $t$,
as mentioned in the above, are typical of the Andreev transport.


\subsection{Effects of the Coulomb interaction in each dot}
\label{subsec:effU}

In order to study the difference between 
the Andreev behavior at $U<2\Gamma_S$ and the Kondo behavior at $U>2\Gamma_S$,
we next consider the case that the Coulomb interaction is switched on 
only in one of the two dots. 
In this situation, the role of the correlation in each dot 
can be seen separately.

\begin{figure}[t] 
\includegraphics[scale=0.73]{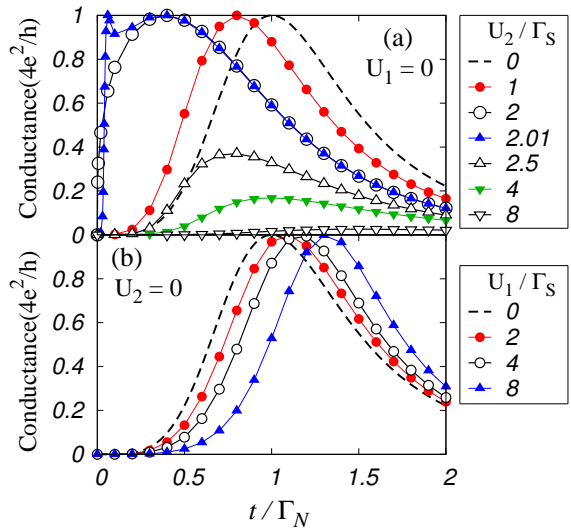}
\caption{
(Color online) 
Conductance for (a) $U_{1}=0$ and (b) $U_{2}=0$
as a function of the inter-dot coupling $t$.
We set $\xi_1=\xi_2=0$ and $\Gamma_N=\Gamma_S$.
}
\label{con_eiU0}
\end{figure}

In Fig.~\ref{con_eiU0} the conductance 
for (a) $U_1=0$ and (b) $U_2=0$ is plotted as a function of $t/\Gamma_N$  
for several values of (a) $U_2$ and (b) $U_1$, respectively.
We can see in Fig.~\ref{con_eiU0}(a) that 
the conductance peak shifts towards the small $t$ region 
as $U_2$ increases from $0$ to $2\Gamma_S$.
Then, for $U_2\gtrsim 2.0 \Gamma_S$ the conductance decreases,
and the peak height becomes smaller than 
the unitary-limit value $4e^2/h$. 
These features are similar to those 
we saw in Fig.~\ref{cont}(a) for $U_1=U_2$.
In contrast  we can see in Fig.~\ref{con_eiU0}(b) 
 that the peak position shifts towards the large $t$ region 
as $U_1$ increases keeping the peak height $4e^2/h$ unchanged.
Since $U_2=0$ in this case, the SC correlation can penetrate
into the QD2 without being disturbed by the Coulomb interaction.  
Therefore, the QD2 can be regarded effectively as a part of the SC host. 
The inter-dot coupling $t$ transmits 
the SC proximity from the QD2 to the QD1.
For this reason, even in the presence of the Coulomb interaction $U_{1}$, 
the conductance can take a maximum with the unitary-limit value $4e^2/h$.  
It explains the reason why the conductance peak shifts 
 towards the larger value of $t$ as $U_{1}$ increases.
We see also in Fig.~\ref{con_eiU0}(b) 
that there are no pronounced qualitative changes 
in the feature of the Andreev transport for large values of $U_1$.
It implies that the effects of $U_1$ 
is mainly to renormalize the inter-dot hopping matrix element $t$.
In contrast, the Coulomb interaction $U_2$ in the dot connected to the SC lead 
causes the qualitative changes, 
particularly near the crossover region between 
the Andreev behavior and Kondo behavior, as seen in Fig.~\ref{con_eiU0}(a).


We next study the precise feature of the crossover, 
choosing the Coulomb interaction at the QD1 to be $U_{1}=0$ 
and $\xi_1=0$, for simplicity.
In this case, the correlation effect of $U_2$ 
on the low-energy properties can be   
described by a local Fermi liquid theory for a single impurity, 
and the fixed-point Hamiltonian 
for free quasi-particles can be expressed in the form,
\cite{Hewson:qp}
\begin{align}
& \widetilde{H}_{qp}^{(0)} \  = \  
-\widetilde{\Delta}_{d2}\left(d_{2\uparrow }^\dag d_{2\downarrow }^\dag
+\textrm{H.c.}\right )
+
\widetilde{t}\, \sum_{\sigma}\left(d_{1\sigma }^\dag d_{2\sigma}^{}
+\textrm{H.c.}\right)
\nonumber\\
& \qquad \quad 
+ \widetilde{\xi}_{2} \left(n_{d,2}-1\right)
+ H_{T,N}  + H_N \;.
\label{Hamiqp}
\end{align}
Here,  
$\widetilde{t}$ and $\widetilde{\Delta}_{d2}$ 
are the renormalized inter-dot coupling 
and the onsite SC potential in the QD2, respectively.
The definitions of these renormalized parameters 
are provided in the appendix \ref{sec:U1eq0}. 
Note that in a special case, for $\xi_2=0$, 
the renormalized level position becomes $\widetilde{\xi}_2 =0$, 
and then the conductance is determined by the ratio of $\widetilde{t}$ 
and $\widetilde{\Gamma}_S$ ($\equiv \widetilde{\Delta}_{d2}$), 
\begin{eqnarray}
G_{U_{1}=0}
=
\frac{4e^2}{h}
\frac
{4\left(\frac{\widetilde{t}^2}{\Gamma_N \widetilde{\Gamma}_S}\right)^2}
{\left\{ 1+\left(\frac{\widetilde{t}^2}{\Gamma_N \widetilde{\Gamma}_S}\right)^2 \right\}^2 }.
\label{eq:Condu-u10}
\end{eqnarray}
The value of these renormalized parameters can be 
deduced from the fixed point of NRG.\cite{Hewson2} 
Particularly  for $\xi_1=\xi_2=0$,
the ratio of the renormalized parameters 
appearing in Eq.\ \eqref{eq:Condu-u10} 
links to the phase shift of 
Bogoliubov particles such that  
$\widetilde{t}^2/(\Gamma_N \widetilde{\Gamma}_S) 
= - \cot \varphi$.

\begin{figure}[t] 
\includegraphics[scale=0.73]{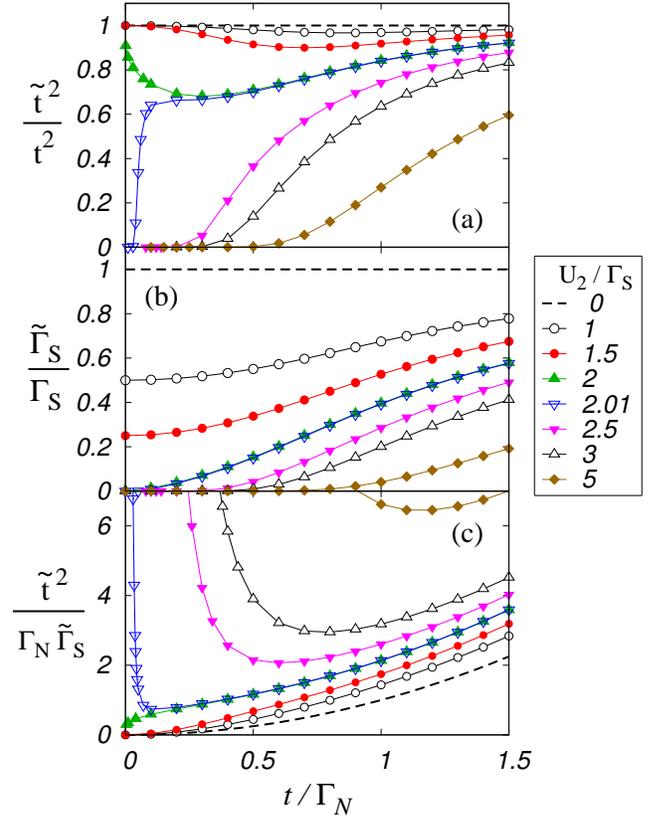}
\caption{
(Color online) 
Renormalized parameters (a) $(\widetilde{t}/t)^2$, 
(b) $\widetilde{\Gamma}_S/\Gamma_S$ and 
(c) the ratio of the parameters 
$\widetilde{t}^2/(\Gamma_N \widetilde{\Gamma}_S)$
as a function of $t/\Gamma_N$
for several values of $U_{2}/\Gamma_S$.
The other parameters are $U_{1}=0$, $\xi_1=\xi_2=0$ and $\Gamma_N=\Gamma_S$.
}
\label{Re-para}
\end{figure}

Figure \ref{Re-para} shows the NRG results of 
$\widetilde{t}$, $\,\widetilde{\Gamma}_S$, and 
the ratio $\widetilde{t}^2/(\Gamma_N \widetilde{\Gamma}_S)$
 as functions of $t/\Gamma_N$ for several values of $U_{2} / \Gamma_S$.
The parameter $(\widetilde{t}/t)^2$ shown in Fig.~\ref{Re-para}(a)
corresponds to the wave function renormalization factor $Z$,
which gives a measure of the correlation effect due to $U_2$
(see also the appendix \ref{sec:U1eq0}).
For weak interactions $U_2 / \Gamma_S \lesssim 1.5$ 
the renormalization factor is almost constant 
$(\widetilde{t}/t)^2 \simeq 1.0$, 
and shows only a weak $t/\Gamma_N$ dependence.
It means that the energy scale of the resonance width is unrenormalized.
We have confirmed that these features are unchanged qualitatively 
in the parameter region of $U_2/\Gamma_S \lesssim 2.0$.
The behavior changes significantly for $U_2 / \Gamma_S \gtrsim 2.0$.
In this large $U_2$ region, 
$(\widetilde{t}/t)^2$ decreases with  $t/\Gamma_N$,
and goes to zero for $t/\Gamma_N \ll 1$.
This indicates that
the system approaches a strongly correlated regime.
In the opposite limit $t/\Gamma_N \gg 1$ 
the renormalization factor gets close to $(\widetilde{t}/t)^2 \to 1$,
and for large $t$ the ground state can be characterized by 
 an inter-dot singlet state. 

The renormalized SC pair potential
$\widetilde{\Gamma}_S$  ($\equiv \widetilde{\Delta}_{d2}$) 
is plotted in Fig.~\ref{Re-para}(b).
We see that $\widetilde{\Gamma}_S$
decreases as the Coulomb interaction $U_2$ increases. 
Furthermore in the limit of $t\to 0$, 
$\,\widetilde{\Gamma}_S$ vanishes 
for strong interactions $U_2 / \Gamma_S \gtrsim 2.0$,
while it reaches a finite value for weak interactions 
 $U_2 / \Gamma_S \lesssim 2.0$.
Reflecting the behavior of $\widetilde{t}$ and $\,\widetilde{\Gamma}_S$, 
the ratio $\widetilde{t}^2/(\Gamma_N \widetilde{\Gamma}_S)$,
which determines the conductance via Eq.\ \eqref{eq:Condu-u10}, 
also shows a clear difference between 
the strong and weak interaction regions.
We see in Fig.~\ref{Re-para}(c) that for $U_2 / \Gamma_S < 2.0$
the ratio $\widetilde{t}^2/(\Gamma_N \widetilde{\Gamma}_S)$ 
decreases monotonically as $t/\Gamma_N$ decreases from $1.5$ to $0$.
In contrast,  $\widetilde{t}^2/(\Gamma_N \widetilde{\Gamma}_S)$ 
takes a minimum for $U_2 / \Gamma_S>2.0$ at an 
intermediate value of $t/\Gamma_N$, 
and then it increases as $t$ decreases further.
\begin{figure}[t] 
\includegraphics[scale=0.73]{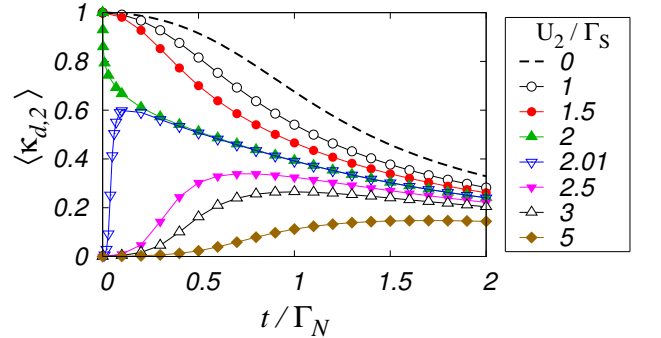}
\caption{
(Color online) 
Pair correlation at the QD2, 
$\langle \kappa_{d,2} \rangle \equiv \langle d_{2\uparrow}^{\dag} 
d_{2\downarrow}^{\dag} + d_{2\downarrow}^{} d_{2\uparrow}^{} \rangle$,
as a function of $t$ for $U_{1}=0$, $\xi_1=\xi_2=0$ and $\Gamma_N=\Gamma_S$.
}
\label{PCQD2}
\end{figure}

The difference between the weak and strong interaction regions
can be seen also in
the expectation values of the pair correlation at the QD2 defined by 
\begin{align} 
\left\langle \kappa_{d,2} \right\rangle  
\,\equiv \,\langle d_{2\uparrow}^{\dag} d_{2\downarrow}^{\dag}
+ d_{2\downarrow}^{} d_{2\uparrow}^{} \rangle
\ = \ 1-\mathcal{Q}_2 \;,
\end{align}
where 
$\mathcal{Q}_2 \equiv 
\langle \widehat{n}_{\gamma, -2} \rangle $ 
is the number of the Bogoliubov particles in the QD2.
Note that in the present case 
the Friedel sum rule described in Eq.\ \eqref{MBogo}
does not give the local charge of a single site,
 so that we have calculated $\mathcal{Q}_2$ directly from the above definition.  Figure \ref{PCQD2} shows the results.
In the weak interaction region $U_2<2\Gamma_S$,
the pair correlation 
$\left\langle \kappa_{d,2} \right\rangle$  
in the QD2 increases as $t$ gets smaller, and
 takes a value $\left\langle \kappa_{d,2} \right\rangle =1$ for $t\to 0$,
which corresponds to 
 the occupation number of the Bogoliubov particle $\mathcal{Q}_2=0$.
In contrast, in the strong interaction region $U_2>2\Gamma_S$, 
$\left\langle \kappa_{d,2} \right\rangle $ has a maximum at finite $t$, 
and then it decreases to reach  $\left\langle \kappa_{d,2} \right\rangle =0$ 
for $t\to 0$, which corresponds to the occupation number $\mathcal{Q}_2=1$.

The significant changes taking place at $U_2 / \Gamma_S = 2.0$ 
 can be explained as a result of the crossover in the ground state. 
In order to figure this crossover out, 
we consider an atomic limit $t \to 0$, 
where the QD2 is connected only to the SC lead.
In this limit the eigenstates of the $H^\mathrm{eff}$ defined 
in Eq.\ \eqref{Hami_seri_eff} can be obtained simply by 
diagonalizing  a single site Hamiltonian of the QD2 
with the onsite SC pair potential $\Delta_{d2} =\Gamma_S$ 
and the Coulomb interaction $U_2$.
Then, the ground state becomes a SC spin-singlet state 
for $U_2<2\Gamma_S$, and a magnetic doublet state for $U_2>2\Gamma_S$.
For small inter-dot coupling $t$, the essential features seen
 in the  atomic limit still remain 
 especially for the local SC singlet state for $U_2<2\Gamma_S$. 
In the doublet region of $U_2>2\Gamma_S$, however, 
the conduction electrons,  
which come from the normal lead through the QD1,
screen the local moment emerging in the QD2 to form a Kondo singlet state.
The level crossing which takes place at $U_2=2\Gamma_S$ in the atomic limit 
becomes the crossover for finite inter-dot coupling.

These two singlet states correspond also to two different fixed points of NRG.
The Hamiltonian $H^\mathrm{eff}$ given in Eq.\ \eqref{eq:after-Bogo} 
describes the interacting Bogoliubov particles, 
as mentioned. In the present case for $\xi_1=\xi_2 =0$, 
the energy level takes the value of $E_2=\Gamma_S$. 
Therefore,  the number of the Bogoliubov particles 
 $\mathcal{Q}_2$ in the QD2 depends crucially on $\Gamma_S$ and $U_2$.
The occupation number approaches the value of
$\mathcal{Q}_2=0$ for small $t$ 
in the weak interaction region $U_2/2<E_2$, and thus 
the QD2 becomes almost empty. 
This  situation corresponds exactly 
to the {\it frozen impurity} fixed point,
which is one of the basic fixed points for the asymmetric Anderson
model.\cite{Wilkins} 
Note that the frozen impurity fixed point 
for the Bogoliubov particles corresponds to 
the local SC singlet state for the original electrons,
as $\left\langle \kappa_{d,2} \right\rangle =1$ for $\mathcal{Q}_2=0$. 
On the other hand, in the strong interaction region of $U_2/2>E_2$, 
the occupation number of the Bogoliubov particles approaches 
 the value of $\mathcal{Q}_2=1$ for small $t\to 0$. 
This situation corresponds to usual Kondo regime,
which is referred to as the {\it strong coupling} fixed point.\cite{Wilkins}

From the above perspective,
we can consider further the behavior of the conductance 
seen in Figs.\ \ref{cont}(a), \ref{contU2} and \ref{con_eiU0}(a).
In the weak interaction region $U_2<2\Gamma_S$, 
the Andreev reflection contributes to the tunneling current. 
This is because the renormalized SC pair potential 
$\widetilde{\Gamma}_S$ ($\equiv\widetilde{\Delta}_{d2}$) 
remains finite in this region, 
although the Coulomb interaction $U_2$ suppresses the SC proximity.
The conductance takes the unitary-limit 
value $4e^2/h$ on the resonance, which 
for $\xi_1=\xi_2 =0$ takes place in the case 
that the ratio $\widetilde{t}^2/(\Gamma_N \widetilde{\Gamma}_S)$ 
becomes equal to $1$. 
The resonance condition can be 
satisfied in the weak interaction region since 
$\widetilde{\Gamma}_S$ and $\widetilde{t}$ are finite. 
The situation is quite different 
in the strong interaction region $U_2>2\Gamma_S$,
where the ground state is a  Kondo singlet. 
 For small inter-dot coupling $t$, 
the renormalized SC pair potential $\widetilde{\Gamma}_S$ 
 is suppressed by the Coulomb interaction $U_2$,
and tends to zero for large $U_2$.
 The renormalization factor $Z$ ($=\widetilde{t}^2/t^2)$ 
also approaches zero in the strong-coupling limit. 
Furthermore, the ratio $\widetilde{t}^2/(\Gamma_N \widetilde{\Gamma}_S)$ 
has a local minimum in the strong coupling case 
as shown in Fig.~\ref{Re-para}(c). 
In a narrow parameter region near the crossover point,
the conductance has the two successive peaks
(see the conductance for $U_2 = 2.01\Gamma_S$
in Fig.\ \ref{con_eiU0}(a)).
This is caused by the fact that the 
value of the ratio $\widetilde{t}^2/(\Gamma_N \widetilde{\Gamma}_S)$ 
at the local minimum becomes less than $1$.
In the case of Fig.\ \ref{con_eiU0}(a),
we have confirmed that the two-peak structure is seen 
in the region  $2\Gamma_S < U_2 < U_2^*$ with $U_2^{*} \simeq 2.05\Gamma_S$. 
For $U_2>U_2^*$, the conductance does not reach the unitary-limit value, 
as the minimum value of the ratio becomes greater than $1$.

\subsection{gate-voltage dependence}

We  have seen  in the above that  
the behavior of the conductance depends strongly on 
which interaction, $U_{1}$ or $U_{2}$, is changed.
In this subsection, we discuss the effects of gate voltage 
on the Andreev transport.
Particularly, we examine the role of 
the energy levels $\varepsilon_1$ in the QD1 
and $\varepsilon_2$ in the QD2 separately, 
choosing the interactions the same $U_1=U_2$.

\subsubsection{$\varepsilon_2$ dependence}

\begin{figure}[t]
\begin{center}
\includegraphics[trim=0mm 0mm 0mm 0mm, scale=0.7]{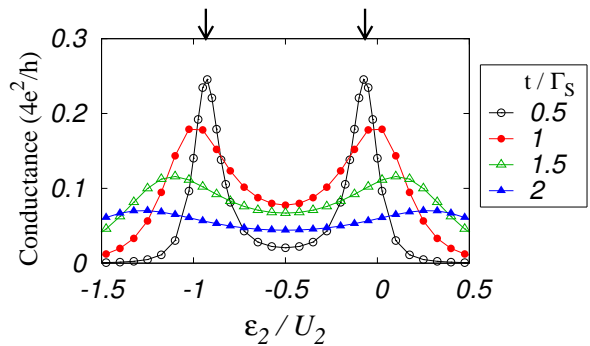}
\end{center}
\vspace{-5mm}
\caption{
(Color online) 
Conductance as a function of the energy level $\varepsilon_2$ in the QD2
for  $\varepsilon_1+U_1/2=0$, $U_1=U_2=4\Gamma_S$ and $\Gamma_N=\Gamma_S$.
The arrows indicate the value of $\varepsilon_2/U_2$ 
at which the singlet-doublet transition occurs 
in the atomic limit $t\to 0$.
}
\label{con-ed2}
\end{figure}

Figure \ref{con-ed2} shows the $\varepsilon_2$ dependence of 
the conductance for several values of $t/\Gamma_S$.
These results are obtained from Eq.~\eqref{eq:Condu},
by choosing the energy level in the QD1 and the other parameters 
such that $\varepsilon_1=-U_1/2$, 
$U_1=U_2 =4\Gamma_S$ and $\Gamma_N=\Gamma_S$.
We see that the conductance has two peaks,
which are broadened as the inter-dot coupling $t$ increases.
These two peaks reflect the level crossing which  takes place  
 at $E_2 =U_2/2$ in the weak coupling limit $t \to 0$.
Here, $E_2 \equiv \sqrt{(\varepsilon_2+U_2/2)^2+\Delta_{d2}^2}$ 
with $\Delta_{d2}\equiv\Gamma_S$ 
is the energy level of the Bogoliubov particles defined 
in Eq.\ \eqref{eq:E_2}. The two arrows in Fig.\ \ref{con-ed2} 
correspond to the critical points, 
and the ground state is a magnetic doublet in the region between the arrows 
$-0.93 \lesssim \varepsilon_2/U_2 \lesssim -0.07$,  
while the ground state becomes a singlet on the outside.

For finite inter-dot coupling $t$, 
the conduction electrons from the normal lead 
contribute to the screening of the local moment in the magnetic doublet state, 
and the ground state for $-0.93 \lesssim \varepsilon_2/U_2 \lesssim -0.07$  
becomes  a Kondo singlet state.\cite{TanakaNQDS}
Therefore, the conductance peaks seen in Fig. \ref{con-ed2} 
are caused essentially  by the crossover between the Kondo singlet state 
and the local SC singlet state, particularly for small $t$. 
As $t$ increases, the ground state varies gradually to 
an inter-dot singlet between the QD1 and QD2.

\subsubsection{$\varepsilon_1$ dependence}

\begin{figure}[t]
\begin{center}
\includegraphics[trim=0mm 0mm 0mm 0mm, scale=0.7]{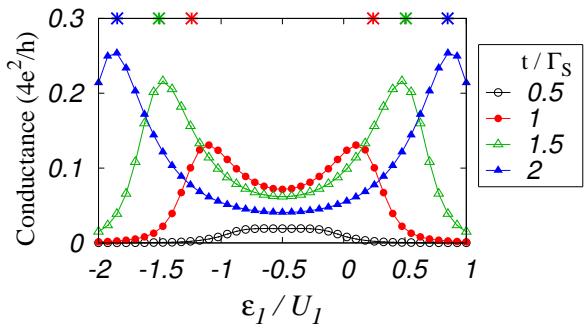}
\end{center}
\vspace{-5mm}
\caption{
(Color online) 
Conductance as a function of the energy level $\varepsilon_1$ in the QD1
for  $\varepsilon_2+U_2/2=0$,  $U_1=U_2= 4 \Gamma_S$ 
and $\Gamma_N=\Gamma_S$.
The symbols ($*$) on an upper horizontal axis correspond to  
the phase boundaries shown in Fig.~\ref{phase-e1}.
}
\label{con-ed1}
\end{figure}

We next consider the $\varepsilon_1$ dependence of the conductance.
To this end, we have to carry out the NRG calculations 
in the situation $\varepsilon_1+U_1/2 \neq 0$ in the QD1 
and $\Delta_{d2}\neq 0$ in the QD2. 
In this case, the system no longer has the uniaxial pseudo-spin symmetry 
mentioned in Sec.\ \ref{subsec:bogo},
and the number of Bogoliubov particles 
$\widehat{{\mathcal Q}}_\mathrm{tot}$ is not conserved.
For this reason, we have calculated 
the conductance using the Kubo formula 
given in Eq.~\eqref{conKubo}.

\begin{figure}[t] 
\includegraphics[scale=0.65]{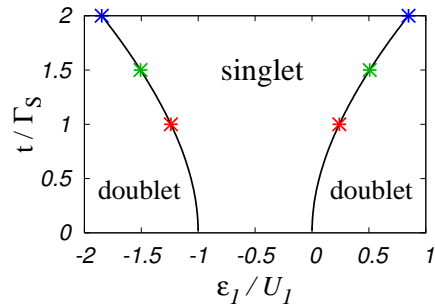}
\caption{
(Color online) 
Phase diagram of the ground state of 
 an isolated DQD system described by $H^\mathrm{eff}_S + H_{DQD}$,
for $U_1=U_2=4\Gamma_S$ and  $\varepsilon_2+U_2/2=0$.
Symbols indicate the points of the single-doublet transition 
for $t/\Gamma_S=1.0, 1.5$ and $2.0$.
}
\label{phase-e1}
\end{figure}

Figure \ref{con-ed1} shows the conductance at zero temperature 
as a function of $\varepsilon_1$ for several values of $t/\Gamma_S$,
choosing the Coulomb interactions 
to be $U_1=U_2=4\Gamma_S$ and $\varepsilon_2=-U_2/2$.
For weak inter-dot coupling of the value of $t/\Gamma_S=0.5$,
the ground state can be characterized by the Kondo singlet  
at $-1.0 \lesssim \varepsilon_1 /U_1\lesssim 0.0$,
and the conductance is suppressed by the Coulomb interaction.
In contrast, for the strong coupling cases  $t/\Gamma_S \gtrsim 1.0$,  
 the ground state near $\varepsilon_1/U_1 \simeq -0.5$ 
corresponds the inter-dot singlet as seen in Fig.\ \ref{cont}.
In this case, the conductance shows two peaks, and 
the peaks move away from the symmetric point 
at $\varepsilon_1/U_1=-0.5$ as $t$ increases.
In order to clarify these features, we have diagonalized 
a two-site Hamiltonian $H^\mathrm{eff}_S + H_{DQD}$ 
consisting of the QD1 and QD2.
The ground-state phase diagram of this cluster   
is plotted in Fig.~\ref{phase-e1} 
in a $\varepsilon_1/U_1$ vs $t/\Gamma_S$ plane.
The ground state is a spin-singlet 
in the central region between the two lines of the phase boundary,
and is a spin-doublet state on the outside.  
Comparing this figure with Fig. \ref{con-ed1},
we see that the positions of the conductance peaks 
agree well with the phase boundary  
between the singlet and doublet ground states of the isolated DQD.

We now consider the screening 
of the local moment emerging at the QD2 
in the doublet regions seen in Fig.~\ref{phase-e1}.
In these regions the energy level of the QD1 is away from the Fermi level, 
and thus the QD1 is almost empty or doubly occupied.
Therefore, the spin degree of freedom is frozen in the QD1, 
and it does not contribute to  the screening. 
Nevertheless, the charge fluctuation still survives in the QD1, 
so that the conduction electrons from the normal lead can tunnel
into the QD2 via the QD1 to screen the local moment. 
This kind of the virtual process
induces an antiferromagnetic exchange coupling between the moment 
in the normal lead and QD2.
This screening process is analogous also to a super-exchange mechanism, 
in which the conduction electrons come over a potential barrier. 
Therefore, the singlet bond which contributes to the Kondo screening 
becomes long in this case. This state has a deformed Kondo cloud, 
and is different from the singlet state that we discussed    
in Sec.\ \ref{subsec:t-dep} and \ref{subsec:effU} 
because near half-filling the spin degree of 
freedom at the QD1 contributes to the screening.
From these observations, it can be concluded that the conductance peaks 
in Fig.~\ref{con-ed1} correspond to   
the crossover between the inter-dot singlet and  
the long-range Kondo singlet.


\section{Summary}

We have studied the Andreev transport through a DQD coupled in series
to normal and SC leads, using the NRG method.
We have demonstrated that 
in the case that the energy level of the QD adjacent to the normal lead is 
on the electron-hole symmetric point, namely for $\xi_1=0$,
the system can be mapped onto a two-impurity Anderson 
model for the interacting Bogoliubov particles.
In this case, the conductance at zero temperature 
is determined by the phase shift of the Bogoliubov particles. 

In order to clarify the ground-state properties,
we have obtained the local SC correlation at the dots and 
the renormalized parameters for the low-energy quasi-particles.
We have shown that 
the low-energy properties 
for a small $t$ region are described by two distinct fixed points
for the Bogoliubov particles.
The local SC singlet state for small interaction  
$U < 2\Gamma_S$ is described by the frozen-impurity fixed point, 
 and the Kondo singlet state for $U > 2\Gamma_S$ is  
characterized by the strong-coupling fixed point.
Moreover, for large $t$, the two dots prefer
to form the inter-dot singlet state.

The conductance in the local SC singlet region for $U < 2\Gamma_S$
takes the unitary-limit value $4e^2/h$ of the Andreev transport.
In the Kondo region, we have found that there is a narrow region 
near the crossover point $U=2\Gamma_S$, in which 
the two-peak structure in the $t$-dependence of the conductance appears.
 Outside this narrow region, the conductance peak in the Kondo regime
gets smaller with increasing $U$.
We have also found that for large $t$, the two dots form
 the inter-dot singlet, and it causes a low conductance in this regime.

We have also calculated the gate-voltage dependence of the conductance,
 by varying the energy level in each dot separately.
The conductance is enhanced 
at the transient region, where the crossover between two different types 
of the singlet states takes place.  
In particular, the energy level $\varepsilon_{1}$ of the QD1 
causes the deformation of the Kondo screening cloud 
and makes the singlet bond long.
Correspondingly, the conductance has the peaks which signify 
the crossover to this long-range Kondo screening. 

The results we have obtained in the present work 
are for the large gap limit $\Delta \to \infty$.
Therefore, the virtual processes using  
the continuum states outside of the SC gap $\Delta$ 
 have not been taken into account.
There are, generally, some quantitative corrections for finite $\Delta$.  
Nevertheless, since the perturbation expansion with respect to $1/\Delta$ 
is applicable to the ground state properties,
the results obtained  in the large gap limit can be 
regarded as the zero-th order contributions.
In fact, in a single QD coupled to a superconductor 
which corresponds to the limit $t \to 0$ of the present system,
the corrections due to finite $\Delta$ vary 
the ground-state phase diagram quantitatively.
Namely, the phase boundary of the singlet-doublet 
transition deviates from the one for the large gap limit $\Delta \to \infty$, 
and the spin-singlet region becomes wide 
for finite $\Delta$.\cite{Oguri,Bauer,Meng}
A similar correction to the crossover energy scale 
will also arise in our system.

Recently, a DQD system coupled to superconductors has been
fabricated using the carbon nanotube \cite{SDQDS_exp}.
The realization of such a system will provide further interesting examples 
of correlation effects in the context of the Andreev transport 
in nanoscale systems.
We hope that the situation we have discussed in this paper 
will be examined experimentally in the near future.

\begin{acknowledgments}
We thank Y.\ Yamada for valuable discussions.
The work was partly supported by a Grant-in-Aid from MEXT Japan
(Grant No.\ 20102008, 21540359).
Y.T.\ was supported previously 
by JSPS Research Fellowships for Young Scientists, 
and currently by Special Postdoctoral Researchers Program of RIKEN.
A.O.\ is supported by JSPS Grant-in-Aid for Scientific Research (C)
(Grant No.\ 20540319).
\end{acknowledgments}

\appendix

\section{Phase shift of the Bogoliubov particles} \label{sec:Deri-con}

We consider the Green's function for the Bogoliubov particles in the DQD,
which is described by the Hamiltonian defined in Eq.~\eqref{eq:after-Bogo} 
for $\xi_1=0$, 
\begin{eqnarray}
\textbf{G}_{DQD}(t-t')
\!\!&=&\!\!
-i\left<
T\left[
\begin{array}{c}
  \gamma_{-1\sigma}^{\phantom{\dagger}} (t)\\
  \gamma_{-2\sigma}^{\phantom{\dagger}} (t)
\end{array} 
 \right]
  \left[ \gamma_{-1\sigma}^{\dagger} (t') , 
  \gamma_{-2\sigma}^{\dagger} (t') \right]
  \right>
\nonumber\\
\!\!&=&\!\!
\left[
   \begin{array}{cc}
    G_{\gamma_{-1} ,\gamma_{-1} } (t-t') & 
    G_{\gamma_{-1} ,\gamma_{-2} } (t-t') \\
    G_{\gamma_{-2} ,\gamma_{-1} } (t-t') & 
    G_{\gamma_{-2} ,\gamma_{-2} } (t-t')
   \end{array}
 \right]_.
\nonumber\\
\label{eq:Green_DQD}
\end{eqnarray}
Note that the matrix form of $\textbf{G}_{DQD}(t-t')$ stems 
not from the Nambu spinor but from a single component Green's function 
for the two interacting dots.
The Fourier transformation of the retarded Green's function for 
Eq. \eqref{eq:Green_DQD}, 
$\textbf{G}_{DQD}^{r} (\varepsilon)$, can be represented as follows,
\begin{eqnarray}
&\left\{
\textbf{G}_{DQD}^{r} (\varepsilon)
\right\}^{-1}
\!\!=\!\!
\left[
 \begin{array}{cc}
 \varepsilon + i\Gamma_N - \Sigma_{11}^{r}(\varepsilon) &
 t-\Sigma_{12}^{r}(\varepsilon) \\
 t-\Sigma_{21}^{r}(\varepsilon) & 
 \varepsilon-E_{2} - \Sigma_{22}^{r}(\varepsilon)
 \end{array}
\right]_,
\nonumber\\
\label{eq:G_DQDinv_re}
\end{eqnarray}
where $\Sigma_{ij}^{r}(\varepsilon)$
 is the self-energy due to the Coulomb interaction.
The advanced Green's function $\textbf{G}_{DQD}^{a} (\varepsilon)$
is also given by taking 
the Hermitian conjugate of $\textbf{G}_{DQD}^{r} (\varepsilon)$,
namely, $\textbf{G}_{DQD}^{a} (\varepsilon) 
=\left\{ \textbf{G}_{DQD}^{r} (\varepsilon) \right\}^{\dag}$.
Because of the time reversal symmetry of the Hamiltonian 
\eqref{eq:after-Bogo}, 
$\Sigma_{12}^{r}(\varepsilon)=\Sigma_{21}^{r}(\varepsilon)$ is satisfied.
At zero temperature, the imaginary part of the self-energy at
the Fermi level $\varepsilon=0$ vanishes, i.e.
${\rm Im}\Sigma_{ij}^{r}(0)=0$.
Therefore, the inverse matrix of $\textbf{G}_{DQD}^r (\varepsilon)$
at $\varepsilon = 0$ can be expressed in the form, 
\begin{eqnarray}
\left\{
\textbf{G}_{DQD}^{r}(0)
\right\}^{-1}
=
\left[
 \begin{array}{cc}
 -\xi^{\star}_{1}+i\Gamma_N &
 t^{\star}  \\
 t^{\star}  & 
 -E^{\star}_{2}
 \end{array}
\right]_,
\label{eq:G_DQDinv_re0}
\end{eqnarray}
where
\begin{eqnarray}
&&
\xi^{\star}_{1}={\rm Re}\,\Sigma_{11}^{r}(0),
\quad
E^{\star}_{2}=E_{2}+{\rm Re}\,\Sigma_{22}^{r}(0),
\nonumber\\
&&\qquad\qquad
t^{\star}=t-{\rm Re}\,\Sigma_{12}^{r}(0).
\label{eq:tilde}
\end{eqnarray}
Using the Friedel sum rule for the Hamiltonian \eqref{eq:after-Bogo},
\cite{Langer}
the local charge at the DQD for the Bogoliubov particles
${\mathcal Q}$ is given by 
\begin{align}
&
{\mathcal Q}
\equiv
\langle \widehat{n}_{\gamma, -2} \rangle 
+ \langle \widehat{n}_{\gamma, -1} \rangle
 = \frac{2 \, \varphi}{\pi}  \;, 
\label{eq:Friedel_app}
\end{align}
where $\varphi$ is the phase shift of the Bogoliubov particles
defined as \cite{Sum:Tanaka}
\begin{align}
\varphi &\equiv
\frac{1}{2{\rm i}} \log\left[
\frac{{\rm det} \left\{ \textbf{G}_{DQD}^{a}(0) \right\}^{-1}}
{{\rm det} \left\{ \textbf{G}_{DQD}^{r}(0) \right\}^{-1}}
\right]
\nonumber\\
&= 
\pi - 
{\rm tan}^{-1}\left(
\frac
{\Gamma_N E_{2}^{\star}}
{(t^{\star})^2-E_{2}^{\star}\xi_{1}^{\star}}
\right)\, . 
\label{eq:phase_shift_app}
\end{align}
The local charge ${\mathcal Q}$ for the Bogoliubov particles 
is proportional to the pair correlation $K$ 
and also to the local charge $M$ for the electrons,    
as mentioned in Sec. \ref{subsec:bogo}.

Similarly, the conductance $G$ in the case of $\xi_1=0$ can be 
expressed in terms of the phase shift $\varphi$.
Using the Landauer formula, the linear conductance
at zero temperature can be expressed 
in the form \cite{TanakaNQDS}
\begin{eqnarray}
G=
\frac{4e^2}{h} \;
4\Gamma_{N}^2 \;
\left|F_{f_{-1} \uparrow ,f_{-1} \downarrow}^r(0)\right|^2.
\label{eq:conV0}
\end{eqnarray}
Here, $F_{f_{-1} \uparrow ,f_{-1} \downarrow}^r(\varepsilon)$
is the Fourier transform of an anomalous Green's function 
for the electrons at the QD1, defined as
$F_{f_{-1} \uparrow ,f_{-1} \downarrow}^r(t-t')=-{\rm i} \Theta(t-t')
\langle \{ f_{-1\uparrow}(t), f_{-1\downarrow}(t') \}\rangle$.
Applying the Bogoliubov transformation of Eq. \eqref{eq:Bogo_B},
$F_{f_{-1} \uparrow ,f_{-1} \downarrow}^r(0)$ can be 
rewritten as
\begin{eqnarray}
F_{ f_{-1} \uparrow ,f_{-1} \downarrow }^r(0)
=
2u v
  {\rm Re}\,
 G^r_{ \gamma_{-1},\gamma_{-1} }(0) \;.
\label{eq:Green-1no12}
\end{eqnarray}
Here, $G_{\gamma_{-1} ,\gamma_{-1} }^r(\varepsilon)$ 
is the Green's function of the Bogoliubov particles at the QD1 
defined in Eq.~\eqref{eq:Green_DQD}
as the (1,1) element of $\textbf{G}_{DQD}^{r}(\varepsilon)$,
and the value at $\varepsilon=0$ is given in Eq.~\eqref{eq:G_DQDinv_re0}. 
Using this value with Eqs.~\eqref{eq:Bogo_factor_B} 
and \eqref{eq:Green-1no12}, 
the formula given in \eqref{eq:conV0} can be expressed in 
terms of the phase shift,   
\begin{eqnarray}
G
\!\!&=&\!\!
\frac{4e^2}{h} 
\left(
\frac
{\Delta_{d2}}{E_{2}}
\right)^2
\frac
{
4 \left(
\frac{
(t^{\star})^2-E_{2}^{\star} \xi_{1}^{\star}
}
{\Gamma_N E_{2}^{\star}}
\right)^2
}
{
\left\{1+
\left(
\frac{
(t^{\star})^2-E_{2}^{\star} \xi_{1}^{\star}
}
{\Gamma_N E_{2}^{\star}}
\right)^2
\right\}^2
}
\nonumber\\
\!\!&=&\!\!
\frac{4e^2}{h} 
\left(
\frac
{\Delta_{d2}}{E_{2}}
\right)^2
\textrm{sin}^2 2\varphi_.
\label{eq:conV0_fin}
\end{eqnarray}
In the second equality, 
we have used the relation Eq. \eqref{eq:phase_shift_app}.

\section{Renormalized parameters for $U_1=0$ and $\xi_1=0$} \label{sec:U1eq0}

In the case the Coulomb interaction in the QD1 is zero $U_1=0$ 
with $\xi_1=0$, 
it is possible to deduce the renormalized parameters 
for the Bogoliubov particles using NRG.
This is because the self energies $\Sigma_{11}^{r}(\varepsilon)$ 
and $\Sigma_{12}^{r}(\varepsilon)$ become zero in this case, 
and the retarded Green's function 
$\textbf{G}_{DQD}^r (\varepsilon)$ in eq. \eqref{eq:G_DQDinv_re} 
takes a simplified form
\begin{eqnarray}
&\left\{
\textbf{G}_{DQD}^r (\varepsilon)
\right\}^{-1}
\!\!=\!\!
\left[
 \begin{array}{cc}
 \varepsilon+i\Gamma_N & t \\
 t & \varepsilon-E_{2} - \Sigma_{22}^{r}(\varepsilon)
 \end{array}
\right]_.
\nonumber\\
\label{eq:G_DQDinv_reU10}
\end{eqnarray}
The asymptotic form of the Green's function at the QD2 
for small $\varepsilon \simeq 0$ at zero temperature is given by
\begin{eqnarray}
&G_{\gamma_{-2}, \gamma_{-2}}^r(\varepsilon)
\, \simeq \,
\frac{\displaystyle \mathstrut Z}{\displaystyle \mathstrut 
\varepsilon- \widetilde{E}_{2}
- \frac{\displaystyle \mathstrut \widetilde{t}^{\; 2 \phantom{|}}}
{\displaystyle \mathstrut  \varepsilon+i\Gamma_N}
}  \;, 
\label{eq:free_qp}
\end{eqnarray}
where
\begin{align}
\widetilde{E}_{2}&\,   \equiv   
Z \, \left[ \,E_2 + \Sigma_{22}^{r}(0) \,\right] , \qquad 
\widetilde{t} \equiv  \sqrt{Z} \, t , 
\\
 \qquad
Z & \,\equiv 
\left(  
1-
\left.\!  
\frac{\partial \Sigma_{22}^{r}(\varepsilon)}{\partial \varepsilon}
\right|_{\varepsilon=0} 
\right)^{-1}.
\label{eq:Z_til} 
\end{align}
The free quasi-particles, which are characterized by 
these renormalized parameters,
 particularly  $\widetilde{E}_{2}$ and $\widetilde{t}$,
can be described by an effective Hamiltonian which 
is given in Eq.~\eqref{Hamiqp} 
in terms of the electrons with the parameters defined by  
\begin{align}
\widetilde{\Delta}_{d2} 
&\,\equiv \, 2uv \widetilde{E}_{2} 
=Z \left[ 1 + \frac{\Sigma_{22}^{r}(0)}{E_2} \right] \, \Delta_{d2}
\;, \\ 
\widetilde{\xi_{2}} 
&\,\equiv \, (u^2-v^2) \widetilde{E}_{2} 
=Z \left[ 1 + \frac{\Sigma_{22}^{r}(0)}{E_2} \right] \, \xi_{2}
\;.
\end{align}
Here, we have used $u$ and $v$ defined in Eq. \eqref{eq:Bogo_factor_B}.


\begin{thebibliography}{99}
\bibitem{condition} 
The large SC gap limit ($\Delta \to \infty$) 
describes reasonably the situation 
where the gap is much greater than the other energy scales, namely for 
$\Delta \gg \max(\Gamma_S, \Gamma_N, t, U_1, U_2, 
\varepsilon_1, \varepsilon_2)$.
\end{thebibliography}
\end{document}